\documentclass[aps,prl,twocolumn,superscriptaddress,showpacs,floatfix,nofootinbib]{revtex4}
\usepackage{amsmath,amssymb,graphicx}
\usepackage{dcolumn} 

\begin{document}
\voffset=0.5 in  

\renewcommand{\topfraction}{0.99}
\renewcommand{\bottomfraction}{0.99}

\renewcommand{\textfraction}{0.01}

\title{
Observation of an Exotic S = $-2$, Q = $-2$ Baryon
Resonance in Proton-Proton Collisions at the CERN SPS}
\affiliation{NIKHEF, Amsterdam, Netherlands.}
\affiliation{Department of Physics, University of Athens, Athens, Greece.}
\affiliation{Comenius University, Bratislava, Slovakia.}
\affiliation{KFKI Research Institute for Particle and Nuclear Physics, Budapest, Hungary.}
\affiliation{MIT, Cambridge, MA, USA.}
\affiliation{Institute of Nuclear Physics, Cracow, Poland.}
\affiliation{Gesellschaft f\"{u}r Schwerionenforschung (GSI), Darmstadt, Germany.}
\affiliation{Joint Institute for Nuclear Research, Dubna, Russia.}
\affiliation{Fachbereich Physik der Universit\"{a}t, Frankfurt, Germany.}
\affiliation{CERN, Geneva, Switzerland.}
\affiliation{University of Houston, Houston, TX, USA.}
\affiliation{\'{S}wietokrzyska Academy, Kielce, Poland.}
\affiliation{Fachbereich Physik der Universit\"{a}t, Marburg, Germany.}
\affiliation{Max-Planck-Institut f\"{u}r Physik, Munich, Germany.}
\affiliation{Institute of Particle and Nuclear Physics, Charles University, Prague, Czech Republic.}
\affiliation{Nuclear Physics Laboratory, University of Washington, Seattle, WA, USA.}
\affiliation{Atomic Physics Department, Sofia University St. Kliment Ohridski, Sofia, Bulgaria.}
\affiliation{Institute for Nuclear Studies, Warsaw, Poland.}
\affiliation{Institute for Experimental Physics, University of Warsaw, Warsaw, Poland.}
\affiliation{Rudjer Boskovic Institute, Zagreb, Croatia.}

\author{C.~Alt}
	\affiliation{Fachbereich Physik der Universit\"{a}t, Frankfurt, Germany.}
\author{T.~Anticic} 
        \affiliation{Rudjer Boskovic Institute, Zagreb, Croatia.} 
\author{B.~Baatar}
	\affiliation{Joint Institute for Nuclear Research, Dubna, Russia.}
\author{D.~Barna} 
        \affiliation{KFKI Research Institute for Particle and Nuclear Physics, Budapest, Hungary.}
\author{J.~Bartke} 
        \affiliation{Institute of Nuclear Physics, Cracow, Poland.} 
\author{M.~Behler} 
	\affiliation{Fachbereich Physik der Universit\"{a}t, Marburg, Germany.}
\author{L.~Betev} 
        \affiliation{CERN, Geneva, Switzerland.}
        \affiliation{Fachbereich Physik der Universit\"{a}t, Frankfurt, Germany.} 
\author{H.~Bia{\l}\-kowska} 
        \affiliation{Institute for Nuclear Studies, Warsaw, Poland.} 
\author{A.~Billmeier} 
        \affiliation{Fachbereich Physik der Universit\"{a}t, Frankfurt, Germany.}
\author{C.~Blume} 
        \affiliation{Gesellschaft f\"{u}r Schwerionenforschung (GSI), Darmstadt, Germany.} 
        \affiliation{Fachbereich Physik der Universit\"{a}t, Frankfurt, Germany.} 
% ...

\collaboration{NA49 Collaboration} \noaffiliation

\date{\today}

\begin{abstract}
Results of resonance searches in the
$\Xi^{-} \pi^{-}$, $\Xi^{-} \pi^{+}$, $\overline{\Xi}^{+}\pi^{-}$ and $\overline{\Xi}^{+}\pi^{+}$ 
invariant mass spectra in proton-proton collisions at $\sqrt{s}=$17.2~GeV are presented.
Evidence is shown for the existence of a narrow $\Xi^{-}\pi^{-}$ baryon resonance 
with mass of 1.862 $\pm$ 0.002 GeV/$c^{2}$ and width below the detector resolution
of about 0.018 GeV/$c^{2}$. The significance is estimated to be 4.0~$\sigma$.
This state is a candidate for the hypothetical exotic $\Xi_{\frac{3}{2}}^{--}$
baryon with S = $-2$, I = $\frac{3}{2}$ and a quark content of ($dsds\bar{u}$).
At the same mass a peak is observed in the $\Xi^{-} \pi^{+}$  spectrum which is a candidate for the $\Xi_{\frac{3}{2}}^{0}$
member of this isospin quartet with a quark content of ($dsus\bar{d}$).
The corresponding antibaryon spectra also show enhancements
at the same invariant mass.
\end{abstract} 


\maketitle

Recent experimental evidence for the first manifestly exotic baryon state 
opens a new chapter in spectroscopy and
may help to elucidate the strong interaction in the strong coupling regime. 
A resonance state was observed \cite{leps,diana,clas,saphir} in the $n K^{+}$ 
and $p K_s^{0}$ invariant mass spectra
near 1.540 GeV/$c^2$ with a width smaller than the experimental resolution of
0.009 GeV/$c^2$. This strangeness S = +1 baryon may be identified with the pentaquark 
state $\Theta^{+}$ with quark content ($udud\bar{s}$). 

Pentaquark states have been theoretically investigated since a long time in
the context of the constituent quark model \cite{jaffe,sorba,strottman,roisnel}.
Some of these are expected to have charge and strangeness quantum number
combinations that cannot exist for three-quark states.
Using the chiral soliton model an anti-decouplet of baryons was predicted by
Chemtob \cite{chemtob}. The lightest member was estimated by
Praszalowicz \cite{praszal} to lie at a mass of 1.530 GeV/$c^2$. 
Diakonov et al. \cite{diakonov} subsequently derived for this
exotic baryon resonance with S = +1, 
J$^{P} = \frac{1}{2}^{+}$ a width of less than 0.015 GeV/$c^2$.  The
mass and width of the experimentally observed $\Theta^{+}$ are close to the theoretical values.
The authors further made predictions for the heavier members of the anti-decuplet,
with the isospin quartet of S = $-2$ baryons having a mass of about 2.070 GeV/$c^2$ and
partial decay width into $\Xi \pi$ of about 0.040 GeV/$c^2$. 
This isospin $\frac{3}{2}$ multiplet contains two $\Xi_{\frac{3}{2}}$ with ordinary charge
assignments ($\Xi_{\frac{3}{2}}^{0},\Xi_{\frac{3}{2}}^{-}$) in addition to the
exotic states $\Xi_{\frac{3}{2}}^{+} (uuss\bar{d})$ and  $\Xi_{\frac{3}{2}}^{--} (ddss\bar{u})$.
The  $\Xi_{\frac{3}{2}}$ isospin quartet  has also been discussed as a part of
higher multiplets (see for example \cite{walliser}).
Jaffe and Wilczek \cite{wilczek} on the other hand base their predictions
on the strong color-spin correlation force and suggest that the $\Theta^{+}(1540)$
baryon is a bound state of two highly correlated $ud$ pairs and an antiquark.
In their model the  $\Theta^{+}(1540)$
has positive parity and lies in an almost ideally mixed $\overline{10}_{f} \oplus 8_{f}$
multiplet of SU(3)$_{f}$.
For the isospin $\frac{3}{2}$ multiplet of $\Xi$s they predict a mass around
1.750 GeV/$c^2$ and a width 50\% greater than that of the $\Theta^{+}(1540)$.
This paper presents the first experimental evidence for the existence of the exotic
$\Xi_{\frac{3}{2}}^{--}$ member of the $\Xi$ multiplet.

Experiments reporting the $\Theta^{+}$ were conducted at energies close to
its production threshold. This paper presents the results of a search
for  the $\Xi_{\frac{3}{2}}^{--}$ and $\Xi_{\frac{3}{2}}^{0}$
states and their antiparticles in proton-proton collisions at $\sqrt{s}=$ 17.2~GeV.
The $p K_{s}^{0}$ decay channel of the $\Theta^{+}$ baryon is
under investigation. However, the combinatorial background is larger
than for the $\Xi_{\frac{3}{2}}$ resonances and no significant signal has been observed yet.

Events were recorded at the CERN SPS
accelerator complex with the NA49 fixed target large acceptance hadron detector~\cite{na49}.
The NA49 tracking system consists of four large volume (50 m$^{3}$) time projection chambers (TPCs).
Two of the TPCs (VTPC1 and VTPC2) are placed inside
superconducting  dipole magnets. Downstream of the magnets  two larger TPCs
(MTPC-R and MTPC-L)  provide acceptance at high momenta. 

The interactions were produced with a beam of 158~GeV/$c$ protons  on a cylindrical
liquid hydrogen target of  20 cm length and 2 cm transverse diameter. The trigger used beam
counters in front of the target, together with an anticoincidence counter further downstream.
The measured trigger cross section was 28.2 mb of which 1 mb was estimated to be elastic
scattering. Thus the detector was sensitive to most of the
inelastic cross section of 31.8~mb ~\cite{pdg}.

The used data sample consists of about 6.5~M events. Reconstruction
started with pattern recognition, momentum fitting and finally formation
of global track candidates (spanning multiple TPCs) of charged particles produced
in the primary interaction and at secondary vertices.
For each event the primary vertex was determined. Events in which
no primary vertex was found were rejected.
To remove non-target interactions 
the reconstructed primary vertex had to lie within $\pm$~9~cm
in the longitudinal ($z$) and within $\pm$~1~cm in the transverse ($x,y$) direction from the center
of the target. These cuts reduced the data sample to 3.75~M events.

Particle identification was
accomplished via measurement of the specific energy loss ($dE/dx$) in the TPCs.
After careful calibration the achieved resolution is 3--6\% depending on the reconstructed
track length ~\cite{na49,lasiuk}. The dependence of the measured $dE/dx$ on velocity was
fitted to a Bethe-Bloch type parametrisation.

The first step in the analysis was the search for $\Lambda$ candidates, which were
then combined with the $\pi^{-}$ to form the 
$\Xi^{-}$ candidates.
Next the $\Xi_{\frac{3}{2}}^{--} (\Xi_{\frac{3}{2}}^{0})$ were searched for in
the $\Xi^{-}\pi^{-} (\Xi^{-}\pi^{+})$ invariant mass spectrum, where the
$\pi^{-} (\pi^{+})$ are primary vertex tracks. An analogous procedure was followed
for the antiparticles.

\begin{figure}[hbt!]
\includegraphics*[width=0.5\textwidth]{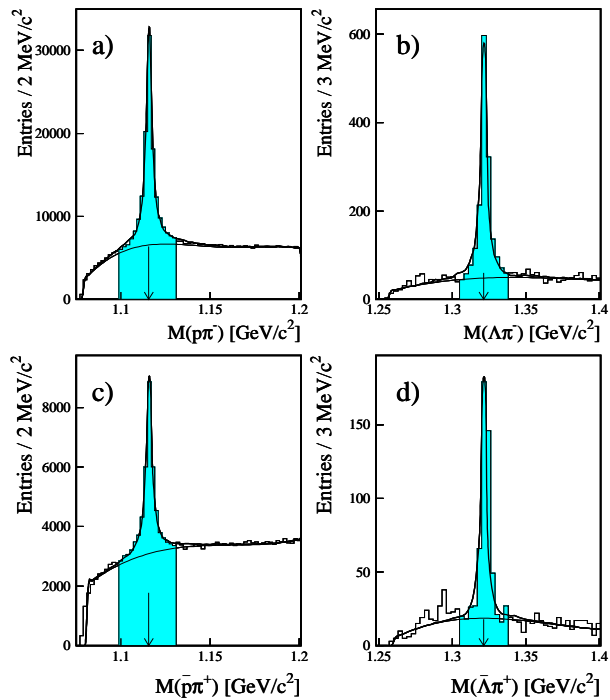}
\caption{\label{fig:lamxi}  
(Color online) 
(a) The $p\pi^{-}$ invariant mass spectrum for V0 topologies. 
(b) The $\Lambda\pi^{-}$ invariant mass spectrum for $\Xi^-$ candidates.
Curves depict the results from a simulation of the detector response,
shaded areas indicate the range of the selected candidates.
The arrows show the nominal $\Lambda$ and $\Xi$ masses.
(c) and (d) show analogous spectra for $\overline{\Lambda}$ and $\overline{\Xi}^{+}$.}
\end{figure}

The protons and pions were selected by requiring  their $dE/dx$ to be within 3~$\sigma$ around the nominal
Bethe-Bloch value.
The $\Lambda$ candidates were identified  by locating the vertices from
neutral decays (so called V0s, mostly upstream of VTPC1). To achieve this, the protons 
were paired with $\pi^{-}$ and both tracked backwards
through the magnetic field. The V0 was constrained to lie on the track with more VTPC points.
A 4-parameter $\chi^{2}$ fit was performed to find the V0 position along the longer track
and the three momentum components of the other track at this point.
The resulting $p\pi^{-}$ invariant mass spectrum is shown in Fig.~\ref{fig:lamxi}a.

To find the $\Xi^{-}$, the $\Lambda$ candidates with a reconstructed
invariant mass within $\pm$~0.015~GeV/$c^{2}$
around the nominal mass (shaded area in Fig.~\ref{fig:lamxi}a)
were combined with all $\pi^{-}$. The fitting procedure
was the same as for V0 finding, but in this case the track parameters
of the $\pi^{-}$ from the $\Xi^{-}$ decay were varied. Several cuts were imposed to increase the significance of the  $\Xi^{-}$ signal.
As the combinatorial background is concentrated close to the primary vertex,
a distance cut of $> 12$ cm  between the primary  and the $\Xi^{-}$  vertex was applied.
Additional cuts on extrapolated track impact positions in the $x$ (magnetic bending) and $y$
(non--bending) directions ($b_{x}$ and $b_{y}$) at the main vertex were imposed.
To ensure that the $\Xi^{-}$ originates from the main vertex its $|b_{x}|$ and $|b_{y}|$ had to
be less than 2~cm and 1~cm, respectively. On the other hand, the $\pi^{-}$ from the
$\Xi^{-}$ decay had to have $|b_{y}|>0.5$~cm.
The resulting $\Lambda\pi^{-}$ invariant mass spectrum is shown in Fig.~\ref{fig:lamxi}b,
where the $\Xi^{-}$ peak is clearly visible.
The $\Xi^{-}$ candidates were selected within
$\pm$~0.015~GeV/$c^{2}$  of the nominal  $\Xi^{-}$ mass.
Only events with one $\Xi^{-}$ candidate (95\%) were retained.
The final data sample used for further analysis consisted of 1640 events containing
one $\Xi^{-}$ and 551 events containing one $\overline{\Xi}^{+}$.

\begin{figure}[hbt!]
\includegraphics*[width=0.50\textwidth]{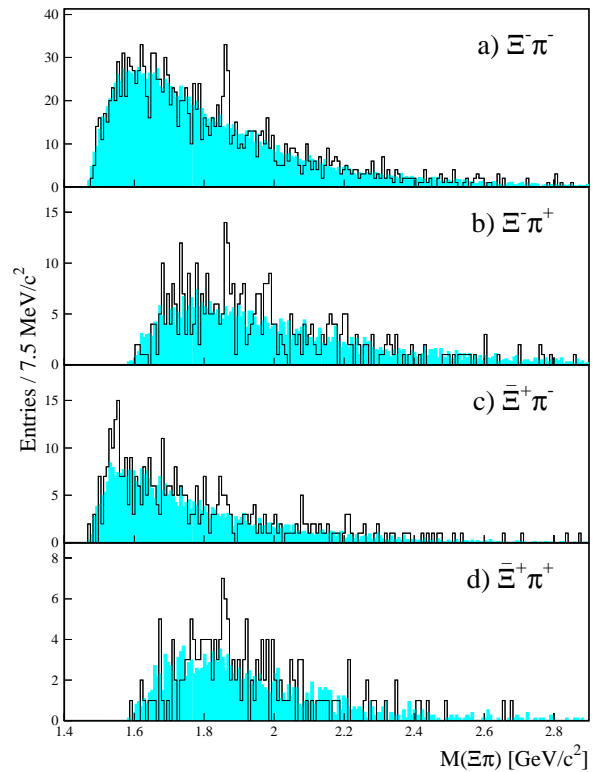}
\caption{\label{fig:pentos-minus}
(Color online)
Invariant mass spectra after selection cuts for $\Xi^{-}\pi^{-}$ (a), $\Xi^{-}\pi^{+}$ (b),
$\overline{\Xi}^{+}\pi^{-}$ (note that the $\overline{\Xi}(1530)^{0}$ state is also visible) (c), 
and $\overline{\Xi}^{+}\pi^{+}$  (d). The shaded histograms are the normalised mixed-event backgrounds.}
\end{figure}

To search for the exotic $\Xi_{\frac{3}{2}}^{--}$ 
the selected $\Xi^{-}$ candidates were combined with primary $\pi^{-}$ tracks.
To select $\pi^{-}$ from the primary vertex, their $|b_{x}|$ and $|b_{y}|$ had to
be less than 1.5~cm and 0.5~cm, respectively, and their $dE/dx$  had to
be within 1.5~$\sigma$ of their nominal Bethe-Bloch value.
Moreover it was found from simulations that the signal to background ratio
in the region of the already visible peak at about 1.86 GeV/$c^{2}$ 
is increased by the restriction $\theta~>~4.5^{o}$ (with $\theta$  the
opening angle between the $\Xi^{-}$ and the $\pi^{-}$ 
in the laboratory frame).
The resulting $\Xi^{-}\pi^{-}$ invariant mass spectrum is shown in
Fig.~\ref{fig:pentos-minus}a. The shaded histogram
is the mixed-event background, obtained by combining the $\Xi^{-}$ and  $\pi^{-}$ from different events and normalising to the number of real combinations. 
A significant narrow peak above the
background is visible at approximately 1.86 GeV/$c^{2}$. The mass window 1.848 - 1.870 GeV/$c^{2}$
contains 81 entries with a background of about $B$ = 45 events. The signal of
$S$ = 36 events has a significance of 4.0 standard deviations  calculated
as $S/\sqrt{S+B}$.
This state is a candidate for the $\Xi_{\frac{3}{2}}^{--}$ pentaquark.

Of the other 3 members of the predicted isospin quartet only the 
$\Xi_{\frac{3}{2}}^{0}$ is
observable in this experiment via the $\Xi^{-}\pi^{+}$ decay channel.
Also the corresponding antibaryon states, $\overline{\Xi}_{\frac{3}{2}}^{++}$  
and $\overline{\Xi}_{\frac{3}{2}}^{0}$, are expected to be produced and should be detectable 
via the $\overline{\Xi}^{+}\pi^{+}$ and $\overline{\Xi}^{+}\pi^{-}$ 
decay channels, respectively.  
In addition to the cuts used for the $\Xi_{\frac{3}{2}}^{--}$ analysis,
a lower cut of 3 GeV/$c^{2}$ was imposed on the $\pi^{+}$ momenta
to minimize the large proton contamination. A lower cut on $dE/dx$ 
of the primary $\pi^{+}$ ($\pi^{-}$) at $-0.5~\sigma$ below the
nominal Bethe-Bloch values in $\Xi^-\pi^+$ ($\overline{\Xi}^+ \pi^-$) 
combinations reduced the kaon contamination.

The mass distributions for $\Xi^{-}\pi^{+}$, $\overline{\Xi}^{+}\pi^{-}$ 
and $\overline{\Xi}^{+}\pi^{+}$
are plotted in Fig.~\ref{fig:pentos-minus}~b,c,d. Indeed, enhancements are seen
in all three spectra. Fits to the combined signal of the $\Xi_{\frac{3}{2}}^{--}$
and its antiparticle and $\Xi_{\frac{3}{2}}^{0}$ and its antiparticle yield peak
positions of 1.862$\pm$0.002 GeV/$c^{2}$ and 1.864$\pm$0.005 GeV/$c^{2}$.
Compared to the $\Xi_{\frac{3}{2}}^{--}$, a smaller rate is expected for the  
$\Xi_{\frac{3}{2}}^{0}$  due to the additional cuts and the 
competing $\Xi^{0}\pi^{0}$ decay channel. Extrapolating
from the yield ratio of about 0.5 between $\overline{\Xi}^{+}$ and $\Xi^{-}$ \cite{tanja} 
one also expects a weaker signal for the $\overline{\Xi}_{\frac{3}{2}}^{++}$  
and $\overline{\Xi}_{\frac{3}{2}}^{0}$.

\begin{figure}[hbt!]
\includegraphics*[width=0.50\textwidth]{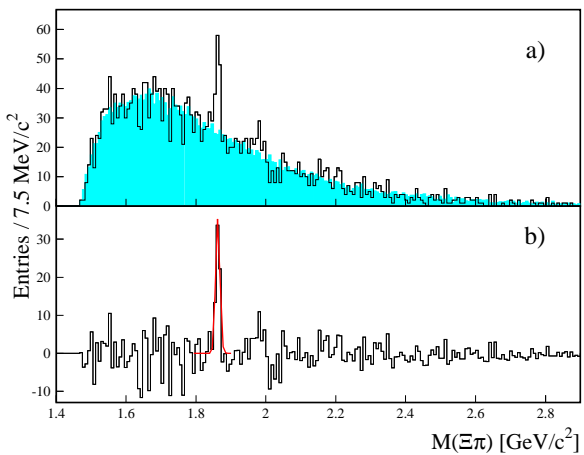}
\caption{\label{fig:pentos-sum}
(Color online)
(a) The sum of the $\Xi^{-}\pi^{-}$, $\Xi^{-}\pi^{+}$, $\overline{\Xi}^{+}\pi^{-}$ and
$\overline{\Xi}^{+}\pi^{+}$  invariant mass spectra.
The shaded histogram shows the normalised mixed-event background.
(b) Background subtracted spectrum with the Gaussian fit to the peak.}
\end{figure}

Finally the sum of the four mass distributions is displayed in Fig.~\ref{fig:pentos-sum}a.
The signal is now $S$ = 67.5 events over a background of $B$ = 76.5, increasing
the significance to 5.6 standard deviations.
Fig.~\ref{fig:pentos-sum}b shows the combinatorial background subtracted distribution.
A Gaussian fit to the peak yields a mass value of 1.862$\pm$0.002~GeV/$c^{2}$
and a FWHM $ = 0.017$~GeV/$c^{2}$ with an error of 0.003 ~GeV/$c^{2}$, largely due to
the uncertainty in the background subtraction. The systematic error on the absolute
mass scale determined from a fit to the $\Xi(1530)^0$ (not shown) is below 0.001~GeV/$c^{2}$.

The detector response to the $\Xi_{\frac{3}{2}}^{--}$ resonance
with  a mass of 1.86 GeV/$c^{2}$ was estimated from simulation. 
The $\Xi_{\frac{3}{2}}^{--}$ was generated with zero mass width, a flat rapidity,
and a thermal transverse momentum distribution with an inverse slope parameter of 160~MeV.
These events were tracked through the detector using GEANT 3.21 followed by a full simulation
of the NA49 apparatus response. They were then reconstructed with the same software as used for
real events. The resulting mass distribution had a $\mbox{FWHM} \approx$ 0.018~GeV/$c^{2}$, 
consistent with the observed width of the $\Xi_{\frac{3}{2}}^{--}$ resonance peak.
The same detector simulation chain was also applied to $\Lambda$ and $\Xi$ production.
The curves in Fig.~\ref{fig:lamxi} demonstrate good agreement with the measured
line shapes and thus confirm the reliability of the simulation. 

The robustness of the $\Xi_{\frac{3}{2}}^{--}$ peak was investigated by varying the $dE/dx$ cut used
for particle selection,
by changing the width of accepted regions around the nominal $\Xi^{-}$ and $\Lambda$
masses, by investigating different event topologies (e.g. the number of $\pi$ mesons per event),
by selecting tracks with different number of clusters, as well as by using different
$b_{x}$ and $b_{y}$ cuts.
Further, the  influence of resonances (including the possibility of particle
misidentification) which could affect the observed peak
was checked by excluding them from the data.
In all cases the peak at 1.86 GeV/$c^{2}$ proved to be robust.
Events generated by the VENUS model \cite{werner} were used to verify that the peak is not
an artifact of the reconstruction.
Finally, a detailed visual inspection of computer displays of the events with $\Xi_{\frac{3}{2}}^{--}$
candidates did not reveal any obvious problem in their quality.

In summary, this analysis provides the first evidence for the existence of a
narrow baryon resonance in the $\Xi^{-}\pi^{-}$ invariant mass
spectrum with a mass of 1.862$\pm$0.002~GeV/$c^{2}$ and a width below the detector resolution
of about 0.018~GeV/$c^{2}$.
The significance is estimated to be about 4.0 $\sigma$.
This state is a candidate for the exotic $\Xi_{\frac{3}{2}}^{--}$
baryon with S = $-2$, I = $\frac{3}{2}$ and a quark content of ($dsds\bar{u}$). Further,
in the $\Xi^{-}\pi^{+}$ invariant mass spectrum
at the same mass an indication is observed of 
the $\Xi_{\frac{3}{2}}^{0}$
member of this isospin quartet with a quark content of ($dsus\bar{d}$).
Also, the corresponding antiparticle spectra show enhancements
at the same invariant mass. Summing the four mass distributions increases the significance 
of the peak to 5.6 $\sigma$.

The observation of the exotic $\Xi_{\frac{3}{2}}^{--}$ 
together with the indication for the $\Xi_{\frac{3}{2}}^{0}$ and their antiparticles
represents an important
step towards experimental confirmation of the existence of the baryon anti-decuplet of
pentaquark states.

\begin{acknowledgments}
This work was supported by the Director, Office of Energy Research,
Division of Nuclear Physics of the Office of High Energy and Nuclear Physics
of the US Department of Energy (DE-ACO3-76SFOOO98 and DE-FG02-91ER40609),
the US National Science Foundation,
the Bundesministerium fur Bildung und Forschung, Germany,
the Alexander von Humboldt Foundation,
the Polish State Committee for Scientific Research (2 P03B 130 23, SPB/CERN/P-03/Dz 446/2002-2004,
 2 P03B 02418, 2 P03B 04123),
the Hungarian Scientific Research Foundation (T032648, T14920 and T32293),
Hungarian National Science Foundation, OTKA, (F034707),
the EC Marie Curie Foundation,
and the Polish-German Foundation.
\end{acknowledgments}

\end{document}